# Observations of photon echo enhancement in an ultraslow light regime

**J. Hahn and B. S. Ham***

*Center for Photon Information Processing, School of Electrical Engineering, Inha University*
*253 Yonghyun-dong, Nam-gu, Incheon 402-751, S. Korea*

Using spectral hole-burning-based ultraslow group velocity in a dilute solid medium, we report enhanced photon echo efficiency three orders of magnitude higher than that in a nonslow light regime. The enhancement is due to exponentially increased absorption of an optical data pulse owing to the enhanced photon-atom interaction in an ultraslow light regime, whereas echo reabsorption is negligibly small due to group-velocity dependent population depletion.

PACS numbers: 42.25.Bs, 42.50.Md, 42.65.-k

Since the first observation of reversible macroscopic atomic coherence [1], photon echoes have been intensively studied for ultrahigh-speed all-optical memories [2-4] as well as multimode quantum memories [5-8]. A major problem of photon echoes is a contradictory effect between absorption of a data pulse and emission of an echo signal resulting in extremely low retrieval efficiency [9]. Because Beer's law determines an exponentially decreased absorption of a traveling light pulse along the propagation direction, absorption of an echo signal in the same direction must be severe due to exponentially increased noninteracted population. Moreover, a critical hurdle for quantum memory applications of photon echoes is population inversion by a $\pi$ rephasing optical pulse resulting in spontaneous emission-triggered echo amplification that violates the no-cloning theorem [10]. Modified photon echo techniques to avoid population inversion have been developed to remove the spontaneous emission noise for a quantum memory protocol [5-8]. A recently proposed double rephasing technique via control deshelving provides a direct solution to the inherent population inversion problem in photon echoes with an additional benefit of storage time extension [11].

Ultraslow light [12] resulting from electromagnetically induced transparency (EIT) (ref. 13) has attracted much attention for quantum memories by mapping a flying quantum bit of light in an atomic ensemble [14-16]. Ultraslow light has also opened a door in low light nonlinear optics [17]. Because ultraslow light enhances photon-atom coupling efficiency, absorption of a traveling light pulse through a dilute medium can also be enhanced. In this Letter, we report slow light-enhanced photon echo efficiency in a dilute medium, with a retrieval efficiency three orders of magnitude higher than that of conventional photon echoes in a nonslow light regime. Unlike EIT-based ultraslow light, the present purpose of using slow light is to enhance absorption of light in a dilute medium, which otherwise is nearly transparent.

Configuration of the present slow-light enhanced photon echo scheme is as simple as with conventional photon echoes, except for the use of a dummy light pulse for slow light preparation. The dummy light, whose frequency is the same as the data pulse, plays a key role in the enhanced photon echo efficiency by burning a spectral hole for group velocity control in an inhomogeneously broadened optical medium. Drawing on the recent observation of dummy light-triggered ultraslow light, the present research benefits ultraslow light applications owing to enhanced coupling efficiency as well as lengthened interaction time using one-color light [18]. We discuss the mechanism of the enhanced photon echoes in terms of propagation distance-dependent absorption strength, where echo signal absorption is significantly alleviated.

Figure 1 shows a schematic of the present ultraslow light-enhanced photon echoes. Figure 1(a) shows a partial energy level diagram of 0.05 at. % $Pr^{3+}$ doped $Y_2SiO_5$ (Pr:YSO) interacting with laser light C1, C2, and P. The light pulses C1 and C2 play a repumping role to transfer atoms from states |1> and |3> to state |2> to increase atom density for P transition. Thus, Fig. 1(a) presents an open two-level system for the transition of P. As shown in Fig. 1(b), the light beams C1 and C2 are set to be noncollinear with P to prevent any unwanted noise from the echo detection. The angle among C1, C2, and P is ~12 mrad and overlapped by 80% through the sample, which is 5 mm in length. As shown in Fig. 1(c), the light P is decomposed into three pulses: H, D, and R. The light H acts as dummy light to prepare the ultraslow light by burning a spectral hole (see ref. 18). Thus, the subsequent data (D) and rephasing (R) pulses experience ultraslow group velocity. The group velocity of D can be easily controlled by adjusting atom population density in the ground state |2> with the power C1, C2, and H as

demonstrated in ref. 18. Here the group velocity of P is inversely proportional to the atom population in the ground state |2>. Under the dummy light H-induced ultraslow light, the data pulse (D) absorption increases significantly due to enhanced photon-atom interactions (will be discussed in Fig. 3). The optical signals detected by an avalanche photodiode are recorded in the oscilloscope by averaging 30 samples. The repetition rate of the light pulse train is 20 Hz. The temperature of the optical medium (Pr:YSO) is kept at ~5 K. The pulse length of C1, C2, and H is 5 ms, 5 ms, and 600 μs, unless otherwise indicated.

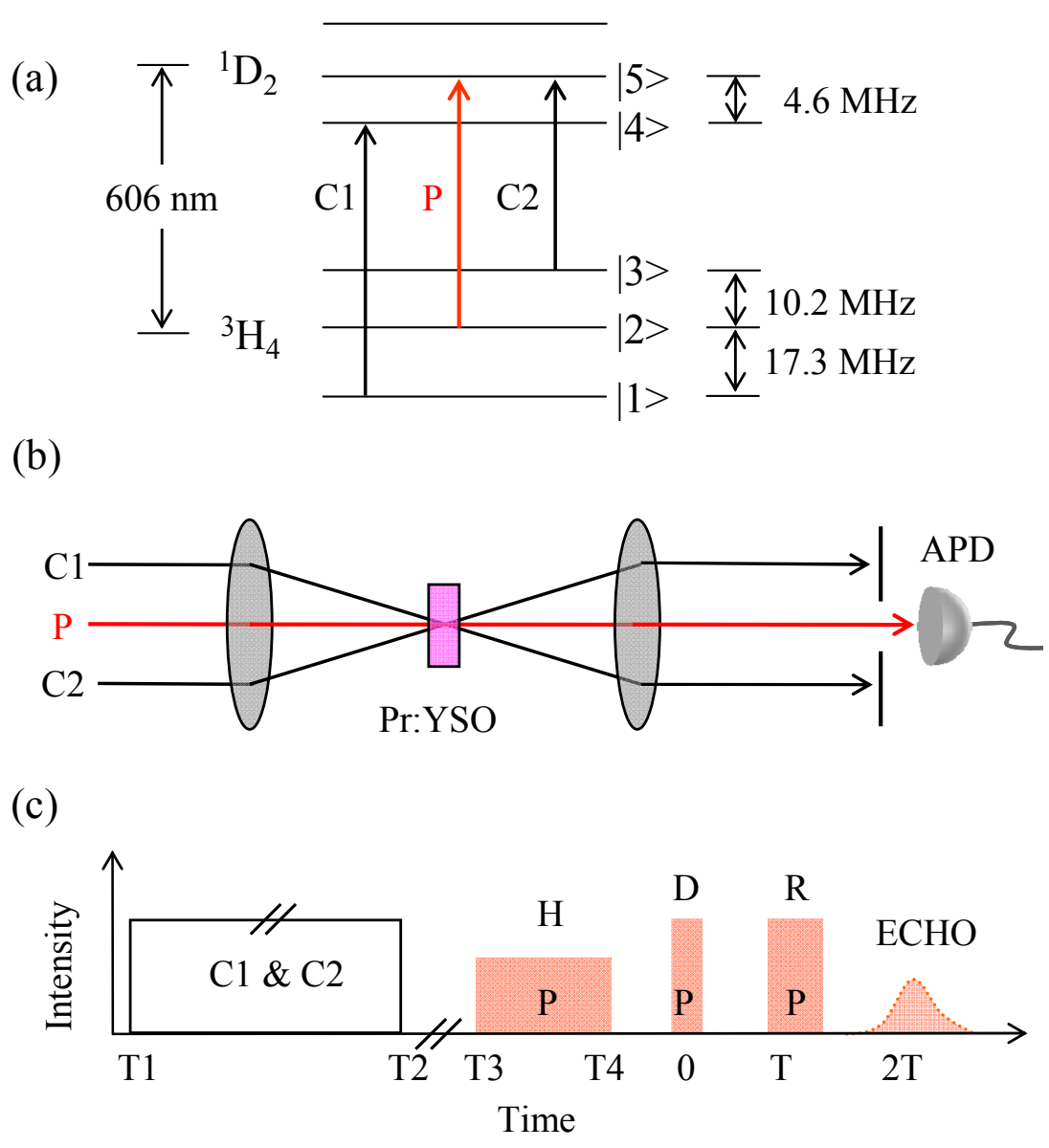

Fig. 1. (Color online) Schematic of ultraslow light-enhanced photon echo. (a) Partial energy level diagram of Pr:YSO interacting with laser light. (b) Pulse propagation. APD: avalanche photo diode, R and C: repump light. P: dummy, data, and rephasing pulse. (c) Pulse sequence. T1=–6.6 ms; T2=–1.6 ms; T3=–1.1 ms; T4=–0.5 ms; T=7 μs. The light is vertically polarized and the crystal axis of Pr:YSO is along the propagation direction **k**.

Figure 2 shows experimental data of the ultraslow light-enhanced photon echoes. The pulse duration of D and R is set at 1.5 μs. Separation between D and R is ~9 μs. The red line in Figs. 2(a) and 2(c) represents input pulse intensity (D and R) before entering the medium, Pr:YSO as a reference. The blue line represents the intensity of the transmitted light without [in Fig. 2(a)] and with [in Fig. 2(c)] the dummy light H. The inset of Fig. 2(a) shows a two-pulse photon echo E in a nonslow light regime. The absorption of the D pulse in Fig. 2(a) is 88.5%, where the corresponding overall optical depth $d$ is calculated by: $d = -\ln\left(\frac{I_{out}}{I_{in}}\right) = -\ln(0.115) = 2.2.$

Figures 2(b) and 2(d) are the integrated pulse area (in intensity) of Figs. 2(a) and 2(c), respectively, where the unit is the reference data (D) pulse area (in intensity). As shown in the inset of Fig. 2(b), the observed photon echo (E) efficiency (intensity ratio of the echo E to the reference D) is as low as 0.7%, which is typical in rare-earth doped solids.

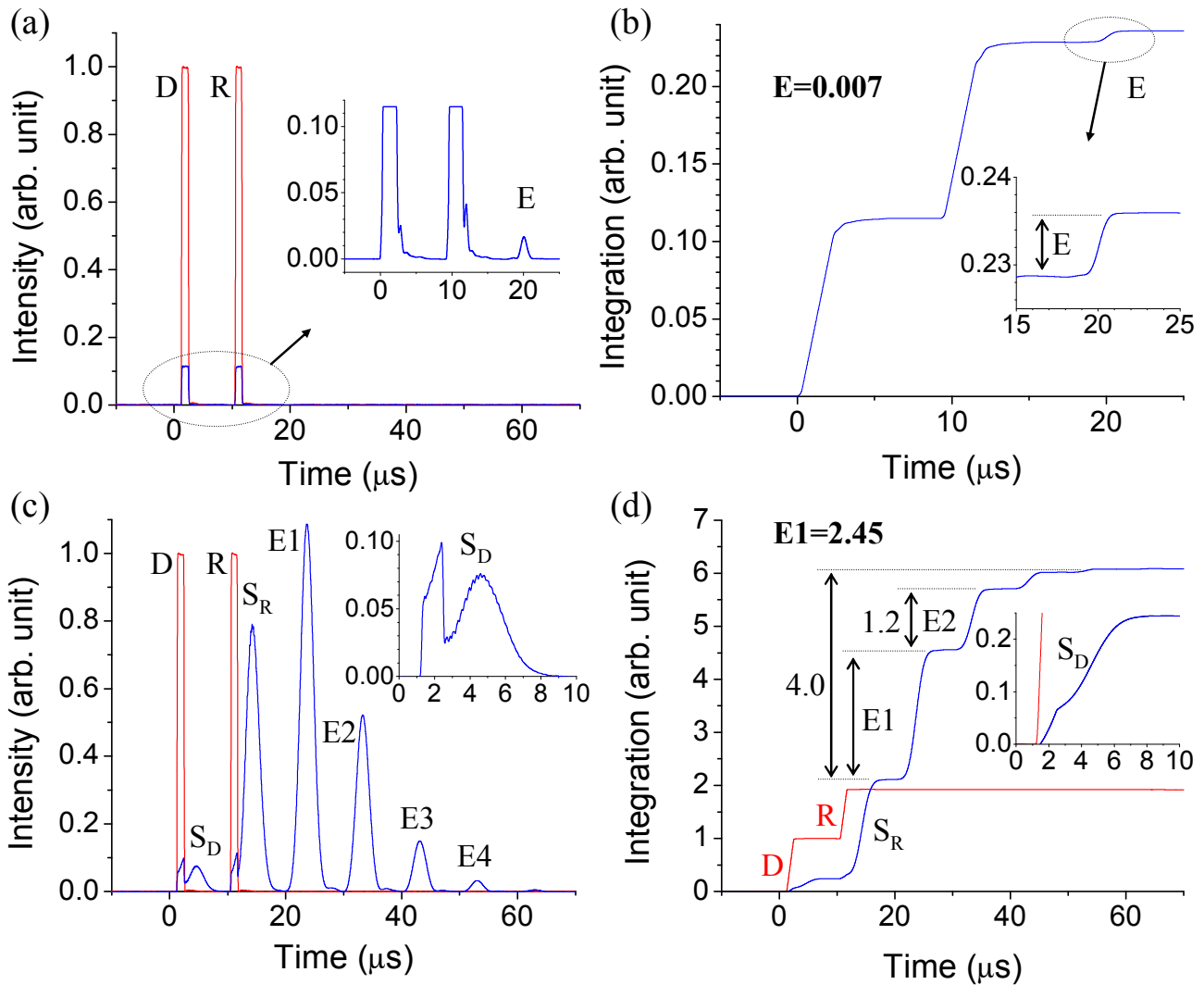

Fig. 2. (Color online) Slow light-enhanced photon echoes. (a) Two-pulse photon echo without H pulse. (b) Pulse integration of (a). The reference is set for the data pulse area in intensity. (c) Two-pulse photon echo with H pulse. $S_D$ and $S_R$ are slow light of D and R, respectively. The echo E2 is from $S_R$ and E1; E3 is from D and E1 as well as E1 and E2. (d) Pulse integration of (c). The reference is set for the data pulse area in intensity. Photon echoes using slow light (red curve) and nonslow light (blue curve). The power of C1, C2, H, and P is 42, 45, 25, and 7 mW, respectively.

Figure 2(c) shows experimental data of the two-pulse photon echoes with the dummy light H. The dummy light H is turned off 500 μs prior to the data pulse D to empty the excited state, where optical population decay time of Pr:YSO is ~160 μs. The inset of Fig. 2(c) shows slow light ($S_D$) of D resulting from the dummy light H-induced spectral hole burning. As shown in the inset of Fig. 2(d), the absorption of the data pulse reaches 76%, which is comparable with that in the nonslow light regime in Fig. 2(a) without the dummy light H. Given that the H pulse depletes a considerable amount of atoms from the ground state |2> as shown in the inset of Fig.

2(c) for rapidly increasing transmission of the nonslow part of D, the comparable absorption in Fig. 2(c) demonstrates the slow light enhanced absorption. With the medium's length of 5 mm and the group delay of 2.6 μs in Fig. 2(c), we calculate that the group velocity $v_g$ of $S_D$ is $v_g \sim 5$ (mm)/2.6 (μs) ~ 2 km/s. Here the group velocity of $S_D$ is controlled by adjusting only the ground state population with H: $N \propto Cos(\sqrt{I_H T_H})$, where $I_H$ and $T_H$ are intensity and pulse duration of H (see Fig. 2 of ref. 18). As a result, amplified photon echoes are observed, where the overall echo amplification is 400%. To the best our knowledge this is the first observation of photon echo amplification without using repumping process [19,20].

As shown in Fig. 2(d), echo E1 efficiency is 245%, and echo E2 efficiency is 120%. Compared with the extremely low photon echo efficiency in the nonslow light regime in Fig. 2(a), the echo amplification in Fig. 2(c) is surprisingly high, with a 350-fold enhancement factor for E1 only. Because data pulse absorption is comparable in both cases, the only source of amplified echoes in the slow light regime is from reduced absorption of echoes (explained below). Although photon echo amplification is beneficial to all-optical signal processing such as associative optical memory [19], the no-cloning theorem prohibits its direct use for quantum memory applications. Combining with a recently proposed double rephasing technique via deshelving for inversionless photon echoes [11], however, the present method also opens a door to quantum memory applications with benefits of near perfect echo efficiency and lengthened photon storage time.

In Fig. 3 we discuss the relationship between slow light and echo intensity. By simply modifying the ground state population with H pulse duration, we obtain a controllable group velocity of $S_D$. Figure 3(a) presents one example of accelerated group velocity, where the absorption of D is decreased to 40% from ~90% in Fig. 2 (see the blue line in the inset). More interestingly, the echo efficiency drops to 20% from 400% (see the red line of the inset), supporting slow-light enhanced photon echoes in Fig. 2. Figure 3(b) represents echo intensity as a function of group velocity (or group delay). The best-fit curve to the data is an exponential function: $A\exp[B(\tau_g - C)]$; $A\exp[B(l/v_g - C)]$, where A, B, and C are arbitrary constants, and $l$ is the medium's length. For Fig. 3(b), A=5, B=3, and C=2.3. Thus, we prove that the echo intensity is exponentially increased as a function of slow factor η, $\eta = c/v_g = (c/l)\tau_g$, where $v_g$ and $\tau_g$ are group velocity and group delay, respectively.

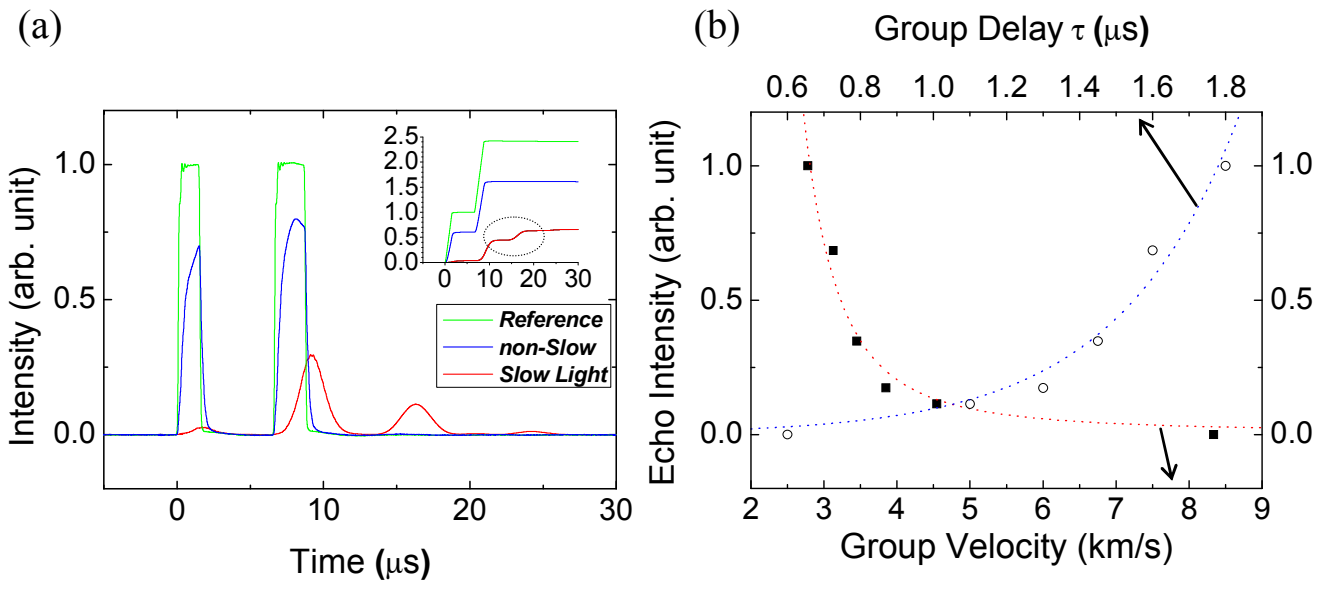

Fig. 3. (Color online) Slow factor dependent echo intensity. (a) Decreased echo efficiency with accelerated group velocity. (b) Echo intensity versus group velocity and group delay. Inset: Integrated pulse area in intensity. The power of P is 11 mW. Echo efficiency calculations are referred to the second bump (dotted circle) of red line in the inset. The pulse duration of R is 1.5 times that of D. Pulse length of DATA and READ is 1.5 μs and 2.3 μs, respectively.

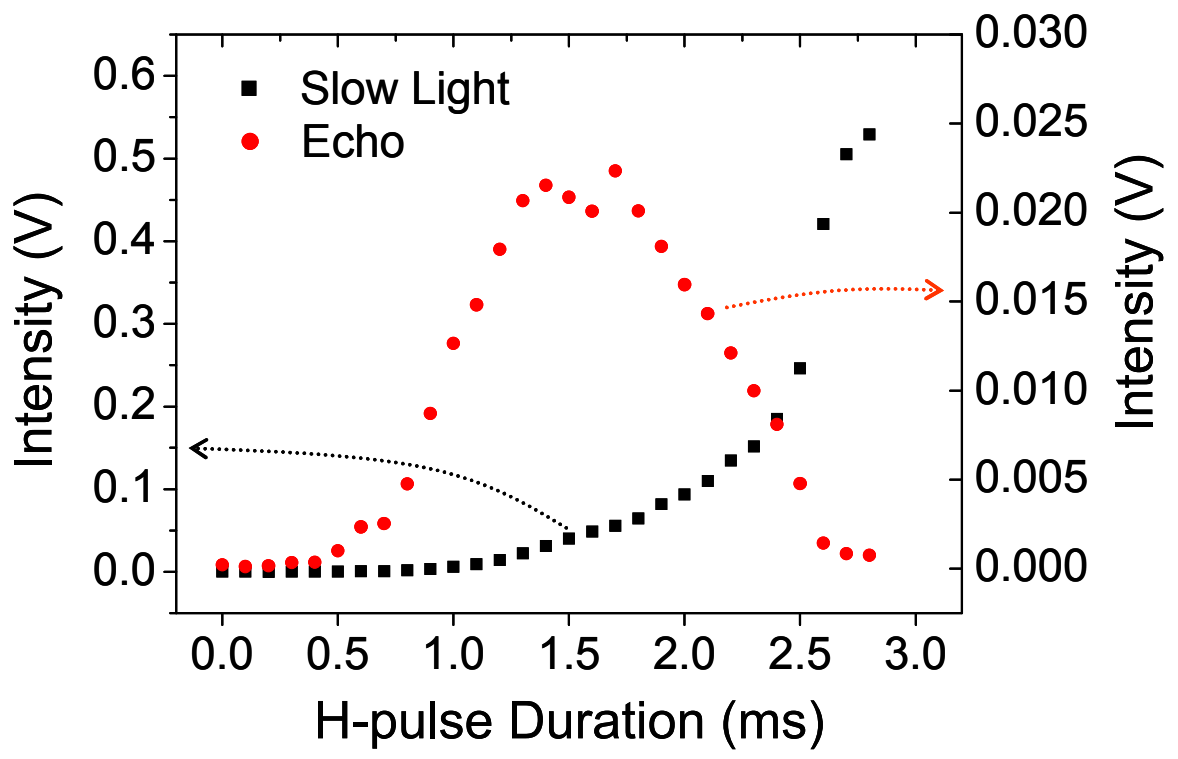

Fig. 4. (Color online) Intensity of photon echo E1 and slow light D versus H-pulse duration.

By Beer's law the ground state population is redistributed to a logarithmic curve as a function of propagation distance by the dummy light H. The position-dependent group delay $\tau_g$ also shows a logarithmic curve by $\tau_g \propto N$ [18]. As shown in Fig. 4, absorption of the data pulse D is exponentially increased with $\tau_g$, resulting in less population along the propagation direction. Thus, the fundamental dilemma between data absorption and echo emission is now removed. This is why the position-dependent group velocity of D becomes a dominating factor over the position-dependent population N in the echo enhancement as shown in Fig. 3(b). In an opposite case of backward dummy light propagation resulting in logarithmic decrease of ground state population, we observed that the photon echo in an ultraslow light regime is weaker than that in a nonslow light regieme (not shown). The peak value of echo intensity in Fig. 4 is due to a mismatch between slow light-enhanced data absorption and the ground state population redistribution based echo emission.

In conclusion we experimentally demonstrated and analyzed ultraslow light-enhanced photon echoes in a rare earth $Pr^{3+}$ doped $Y_2SiO_5$. The observed echo efficiency is three orders of magnitude higher than that of conventional two-pulse photon echoes in a nonslow light regime. Compared with other modified photon echo methods, the present technique is simpler and keeps the same benefit of photon echoes: multidimensional, ultrafast all-optical (quantum) memories.

This work was supported by the CRI program (No. 2010-0000690) of the Korean government (MEST) via National Research Foundation.

[*] *Electronic address: bham@inha.ac.kr*